\documentclass[aip,rsi,preprint]{revtex4-1}

\usepackage{graphicx}% Include figure files

\begin{document}

\title{Stability and Performance of a New Toroidal Laboratory \
Magnetized Plasma Device with Sheared Magnetic Field Lines\
using an Internal Ring Conductor
 }
\author{Th. Pierre} 

\affiliation{Centre National de la Recherche Scientifique, UMR 7345 
Laboratoire PIIM, Aix-Marseille Universit\'e, Marseille, France}

\normalfont\large

\begin{abstract}

In a new toroidal laboratory plasma device including a poloidal magnetic field created by an internal
circular conductor, the confinement efficiency of the magnetized plasma and the turbulence level are
studied in different situations. The plasma density is greatly enhanced when a sufficient poloidal 
magnetic field is established. Moreover, the instabilities and the turbulence usually found in toroidal 
devices without shear of the magnetic field lines are suppressed when the rotational transform is present. 
The measurement of the plasma decay time allows to estimate the confinement time 
of the particles which is compared to the Bohm diffusion time and to the value predicted by different diffusion models, 
especially the neoclassical diffusion involving trapped particles.

\end{abstract}

 \maketitle

\section{INTRODUCTION}

The efficient confinement of plasmas using a magnetic field in closed devices has been 
a major goal in plasma physics during the last 60 years. 

 Unfortunately, 
the basic mechanism of electrons and ions drift inside a device 
exhibiting curvature of the field lines and gradient of the magnetic 
strength leads to a very rapid escape of the particles unless a convenient 
topology of the magnetic field lines is established for a better confinement.

It has been understood very early that only a twist of the field lines can 
cancel the drift of the charged particles. In a twisted configuration of 
the magnetic confinement, the charged particles experience half of the time 
a confining drift and part of the time a deconfining drift along their 
trajectory. External windings have been employed leading to the 
stellarator-type devices for fusion experiments. Internal windings creating a 
helicoidal field has also been used in fusion-oriented devices in the late 
sixties. Finally, the tokamak device with a large current flowing inside 
the plasma itself was shown to be the most efficient device for 
the achievement of thermonuclear fusion. Is is worth noting that the 
confinement of a hot plasma by a dipolar magnetic field,
as it is the case in the new device described here, has been studied
very early in fusion research. For instance, the F-IV device designed by B. Lehnert
has been operated between 1965 and 1970 \cite{lehnert}. In this experiment, a very thick copper coil
was inserted inside a magnetized torus and the quality of the confinement was studied.
A recent similar program has been conducted using a superconducting ring \cite{ldx1}
(Levitated Dipole Experiment, LDX) with the goal to develop a device with a better confinement, 
compared to modern very large tokamaks. However, these devices are not convenient for 
laboratory studies. For instance, plasma diagnostics
are difficult to implement inside a hot and dense plasma created by injection of several
tens of kilowatts of electric power.

On the other hand, basic plasma physics has been investigated inside Simple 
Magnetized Tori (SMT) especially for turbulence studies (ACT-1 \cite{act}, BETA \cite{beta}, 
BLAAMAN \cite{blaaman}, THORELLO \cite{thorello}, TORIX \cite{torix}, TORPEX \cite{torpex}). 
In order to counterbalance 
the destabilizing effect of the curvature of the field line inside an SMT, a vertical magnetic field 
is often applied for a better definition of the connection length along the
B-field line (i.e. the distance between any point inside the plasma 
and the collecting limiter or the wall), inducing a partial control of the turbulence regime.
However, this arrangement cannot pertain for a good confinement of a plasma inside a 
magnetized torus.

\section{The toroidal device}

 We report here an elegant technical arrangement modifying a SMT in 
order to create and to control the shear of the magnetic field lines inside 
a toroidal magnetized plasma. This modification leads to a dramatic change in the confinement
properties of the device.
It is shown that a laboratory toroidal plasma with twisted 
field lines exhibits a large increase of the confinement time compared to a classical
SMT and that consequently an unprecedented high density plasma can be created with  
low injected power.

The new device (MISTOR) consists in a stainless steel toroidal vessel (major radius 
R = 0.6 m, minor radius a = 0.2 m) pumped to a base pressure of $10^{-5}$ Pa and immersed
in a toroidal magnetic field created by a set of 55 water-cooled coils. The large number of coils
insures a very low ripple of the field lines around the torus, which is important
to prevent the development of instabilities induced by the plasma inhomogeneities
at the edge of the torus and by the magnetic trapping of particles inside local magnetic cusps. 
 During the measurements described in this paper, the typical B-field stength on the 
secondary axis is 0.015 Tesla, with a maximum value 0.03 Tesla. 
The Helium working pressure is in the range 0.05 to 0.3 Pa. The plasma 
is created by electrical discharge using only one 
small bended tungsten filament (0.2 mm in diameter, 4 cm in length) located close to the central conductor 
and heated at 2000 K using two insulated electrical connections of small diameter (3 mm). The discharge is created 
between the cathode (heated filament) and the central toroidal conductor used as the anode. 
The central circular conductor can be either grounded or biased. A small positive or negative 
bias compared to ground gives the possibility to control the radial electric field, especially at the edge 
of the toroidal magnetized plasma. The power injected inside the 
torus for plasma creation is low in this new device. The injected power
is considerably lower than the power usually injected in a classical SMT, 
that power ranging typically from several hundred watts to two kilowatts.
The toroidal conductor is made of three superposed turns of a flat copper bar (5mm x 15mm) 
assembled in a series of connected copper arcs whose curvature radius is 60 cm. The median turn 
is insulated by a woven silica sleeve. The section of the internal ring conductor is about 1.6 cm x 1.6 cm . 

A larger number of turns would allow a 
minimization of the magnetic disturbance induced by the electrical feeding of the conductor but the increased resistance 
would induce a larger heat load inside the conductor. The conductor is maintained on the 
secondary axis of the torus using six vertically adjustable holders connected to six 
insulated stainless steel wires (0.5 mm in diameter). The insulation of the holders is 
important because the objects intercepting the magnetic field lines have to be kept
floating to avoid a reduction in plasma density by collection of the ionizing electrons.
The magnetic disturbance arising from the currents feedings of the central conductor is reduced
using two flat conductors very close eachother and parallel to the field 
lines: the electrical feeding is designed with reduced section (2mm width and 
20 mm wide) positioned in the equatorial plane. In typical conditions with 300 A flowing 
in each turn of the conductor, the applied voltage is 1.3 V. 
The dissipated power is about 400 W and the temperature of the conductor after one 
hour operation of the device is about 150$^\circ$ Celsius.

The generated poloidal magnetic field is calculated using the classical evaluation of
the B-field around a magnetic loop (calculation of the exact near field). 
The small aspect ratio of the torus implies that the magnetic surfaces are not exactly toroidal. 
An asymmetry between the high field side and the low field side is present and this has
to be taken into account during the measurements. For instance, choosing a toroidal magnetic field
of 0.015 T and a total toroidal current of 450 A, a point located 10 cm outside in the
equatorial plane (low-field side) is connected to a point located at r = -7 cm 
in the high field side of the torus.
The radial profile of the pitch angle of the field line is an important parameter. It is described by the
safety factor q in the terminology of fusion devices due to the Kruskal-Shafranov limit for
the kink instability at the external edge of the plasma in a tokamak.
At a given radial position, the parameter q is the number of turns a field line
orbits around the principal axis of the torus for one complete orbit around the minor axis.
Given the toroidal magnetic field $B_{T}$ and poloidal field Bp, at radial position R (major radius)
and minor radius position r, q=(r/R)$(B_{T}/B_{p})$. Figure 1 displays the
evolution of the safety factor q across a section of the torus. 

 \begin{figure}
	\includegraphics [scale=0.3]{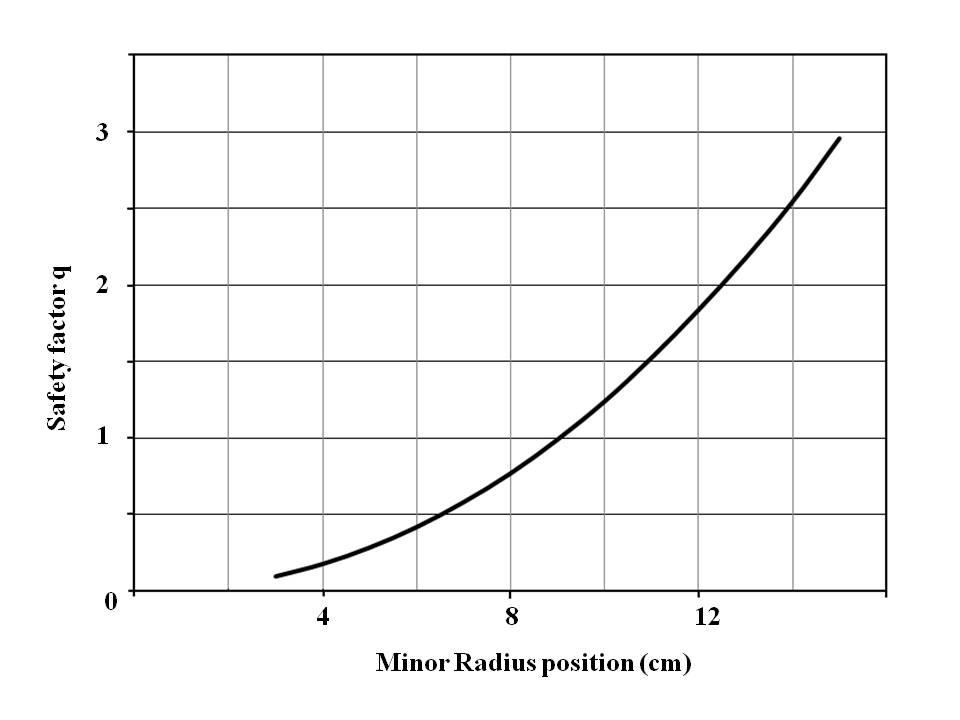}
	\caption{\label{fig:mistor-fig1} Radial evolution of the security factor q
	in typical conditions: toroidal field $B_T$ = 0.015 T on the secondary axis and toroidal current I=750 A). A large
	rotational transform -corresponding to q lower than 1 - is obtained at a radius lower than 9 cm.	
	}
	\end{figure}
	
For typical conditions, with 750 A in the toroidal conductor and with a toroidal B-field strentgh of 0.015 T,
the q value at the plasma edge is larger than 3 and for radii
lower than r = 9 cm, the safety factor is below q=1.
This means that close to the internal conductor, the field lines make several turns around 
the secondary axis during one turn around the principal axis.
As will be seen below, this strong shear of the field lines in typical conditions
induces a good homogeneity and a high density of the plasma.

\section{PLASMA PARAMETERS}

The density and temperature of the electrons are analyzed using several 
radially movable Langmuir probes
located around the torus. At working pressure 0.1 Pa in Helium, the plasma is created
with a discharge voltage of 60 V to 80 V and a discharge current lower than 200 mA. The 
obtained density is in the range $10^{15}$ to $10^{17} m^{-3}$, with a very rapid increase of the density
when the emission current of the heated filament is enhanced.
It is important to note that this relatively high plasma density is obtained with a low injected power.
In this toroidal device, the heating of the cathode is below 40 W and the discharge corresponds to an electric power
lower less 100 W. This can be compared to the power injected for plasma creation in a comparable
linear device (MISTRAL device \cite{pierre}) with the same magnetized plasma diameter,
the same magnetic strength, but smaller plasma volume. 
In this linear device, the heating power is about 2500 W
and the discharge power for creation of a Helium plasma of density $10^{16} m^{-3}$
is close to 500 W. In conclusion, the toroidal device with reduced end-losses is indeed an
efficient way to produce a dense magnetized plasma. 
As will be seen below, the density and the stability
of the plasma can be largely enhanced by implementing the poloidal confinement.
In the toroidal MISTOR device, the electron temperature ranges from 4 to 6 eV 
depending on the working Helium pressure ranging from 0.1 to 0.3 Pa. The ion temperature
is supposed to be lower than 0.1 eV since no heating mechanism of the ions seems to be present.
Helium gas is used in order to keep the ion Larmor radius as small as possible (below 3 mm).
On the other hand, the high ionization energy of Helium leads to a rather high discharge voltage. 
Considering all these parameters, the $\rho$* parameter
(ratio of the ion Larmor radius to the radius of the plasma)is about $1.10^{-2}$. 
This parameter is often used for comparison between various physical situations in 
magnetically confined plasmas in fusion devices.
Assumig a maxwellian distribution, the plasma potential is estimated from the probe voltage at which the evolution of 
the electronic current collected by the biased probe is no longer exponential.
The radial profile of the plasma potential gives the radial electric field.
This is a crucial parameter for the growth rate of drift instabilities.
Moreover, the discharge voltage can be switched-off very quickly using a fast power transistor. In this way,
it is possible to measure the decay time of the plasma density and to investigate
the evolution of the efficiency of the confinement when the poloidal magnetic field is increased. 

\section{OPERATION WITH POLOIDAL B-FIELD}
Numerous studies in SMT devices have shown during the last 
thirty years that these devices are intrinsically turbulent.
This situation has been explained invoking various types of instabilities arising in
different situations, especially the high or low magnetization of the ions, 
the presence of a high or low radial electric field,
or the specific ionization arrangement, for instance thermionic discharge or UHF creation of the plasma.
These instabilities include gradient drift waves,
ExB drift instability, interchange instability, drift interchange instability, Simon-Hoh instability i.e.
the collisionnal slow-ion-drift instability \cite{simon, hoh, kaur}.
The simple magnetized plasma torus is a good test bench for the study of plasma turbulence.
	 
The mean plasma density and the averaged radial electric field are measured when 
no toroidal current is present (SMT configuration) and when the toroidal current flowing in the 
internal circular conductor is progressively increased. 
Using a radially movable plane Langmuir probe whose collecting area is oriented facing the field lines, 
the density profile across a poloidal section of the plasma torus and the averaged electric field deduced from the
radial profile of the plasma potential are compared in two different situations.
The reference experiment is conducted with no toroidal current in the circular conductor.
It is important to note that no vertical magnetic field is applied in this device.
As shown if Figure 2, the plasma exhibits in this case a radial density profile (triangles, dotted line) 
with a steep radial decrease, an averaged plasma potential profile corresponding to a radial electric 
field in the range 100 to 200 V/m, and a very large level of density fluctuations ($\delta$n/n) 
across the whole plasma.

\begin{figure}
	\includegraphics [scale=0.4] {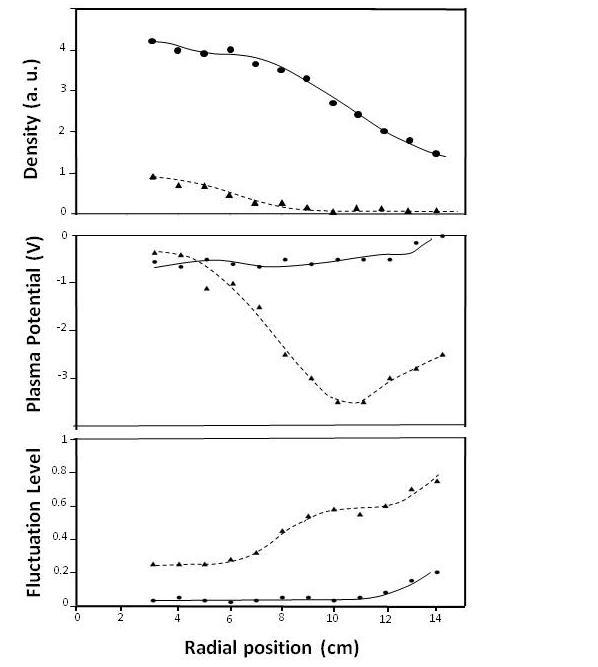}
	\caption{\label{fig:mistor-fig2} Radial profile of the density, the plasma potential, 
and the fluctuation level in typical conditions ($B_T$ = 0.015 T) : when no poloidal field is present (triangles, dotted lines) and 
with a security factor q = 1 at radius r = 8 cm (dots, solid lines). In this latter case, the density is strongly enhanced and the fluation level is largely reduced.
	}
	\end{figure}

On the other hand, when q = 1 is established at radius r = 8 cm, 
as shown in Fig. 2 (dots, solid line), 
the density is higher with a lower density gradient. The potential profile is flatter, with 
a low radial electric field inside the plasma, except at the very edge, and the turbulence 
level is markedly reduced. Only at the edge layer of the plasma, the fluctuation level
increases. This is correlated to a local increase of the radial electric field.
In fact, the turbulence level in the central part of the plasma is typically reduced by more than a factor 10 when the
B-field lines are strongly twisted. In this situation, the ionization is more homogeneous across 
the poloidal section and this contributes to a low radial electric field inside the plasma torus.
This indicates that the observed suppression of the low-frequency instabilities is clearly correlated 
to the vanishing radial electric field. It is suspected that the anomalous transport associated 
with the turbulence is suppressed in this situation, allowing 
a stable and quiet plasma to be created.

The quality of the confinement is investigated by measuring the decay time of the plasma density 
after the discharge voltage is switched-off (fast electronic switch).
At location r = 8 cm in the equatorial plane the low field side, the decay 
time is recorded in two typical situations: without toroidal current and with a security factor 
q = 1 established at radius r = 8 cm. 
In the first case, a turbulent decay of the plasma is recorded 
with a decay time about 100 $\mu$s, as shown in Figure 3, curve (a). 
In the sheared field situation depicted in Figure 3 curve (b), the decay time is much longer.
The decay curve exhibits three different phases that are more 
accurately analyzed with a log-lin plot of the data. During the 
first 50 $\mu$s, a rapid decrease of the density is recorded 
and a second phase is established over the next 400 $\mu$s with a slightly slower decrease of the density 
and a decay time of 0.4 ms. Finally, an exponential decay is present with a longer decay time of 1.8 ms. 
The exact temporal decay law during the first phase is difficult to characterize as an exponential 
decay in a diffusing plasma or as a reciprocal decay in a recombining plasma. 

Further detailed measurements will allow to decide about the mechanisms of the plasma density decay , 
for instance electron-ion recombination, loss of particles on collecting surfaces, or radial diffusion 
of the particles. In the latter case, it is important to investigate wether classical diffusion, Bohm diffusion, 
or neoclassical diffusion is the leading mechanism \cite{maggs1,simon1,maggs2,simon2,fruchtman}. 
A rough estimate of the classical diffusion time can be obtained using typical value of the 
collision times and typical plasma parameters. In the sheared field lines configuration with 
vanishing electric field, the decay can be computed assuming a decay due
to the sole radial diffusion without convection radial velocity. In this device, the electron temperature 
is low (4 eV) compared to the ionization potential of Helium
so a source term for plasma sustainment during the plasma decay is not considered here.
For simplicity, it is possible to use
a model of plasma decay inside a cylinder of radius r = 20 cm with the assumption that recombination 
on the central conductor is negligible. The transport coefficient of the electrons
will determine the decay of the plasma if ambipolar diffusion is assumed. This is a correct hypothesis only if
no conductive surface is collecting the charges and cancelling the radial electric field induced by the difference in
perpendicular mobility of ions and electrons (the so-called Simon short-circuit effect \cite{simon3}).
Assuming the decay inside a cylinder of radius 20 cm, this gives a measured value $D_{exp.}= 9 m^2/s$.

\begin{figure}
	\includegraphics [scale=0.4]{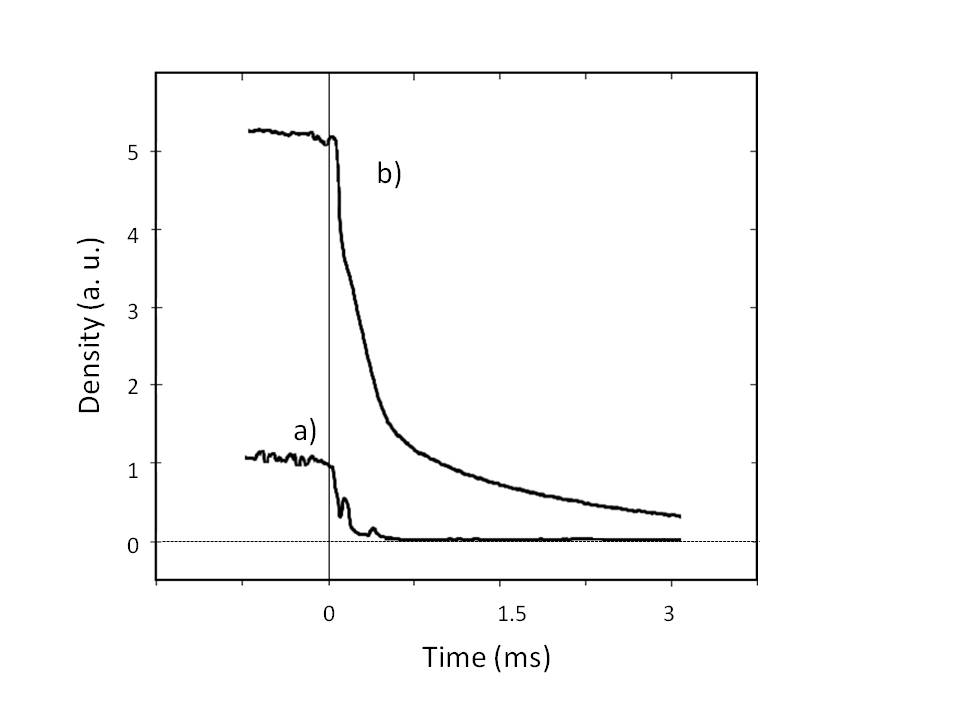}
	\caption{\label{fig:mistor-fig3} Decay of the plasma ($B_T$ = 0.015 T) at radial location r = 4 cm when no toroidal current is present (trace a) and when 800 A are flowing in the circular conductor establishing a 
security factor q = 1 at radius r = 8 cm (trace b).	A strong enhancement of the confinement 
time is recorded in this latter case. 
	}
	\end{figure}

To compare with the experimental value, the diffusion coefficient is calculated
by evaluating the random-walk process inducing the diffusion using an elementary step equal to the electron larmor
radius and a time interval equal to the electron-neutral collision time.
%   $D_{\perp e} = \rho_{e}^2 / 4 \tau_{e-n}$. 
Considering the typical plasma parameters and the Helium working pressure, this transport 
coefficient for the electrons is close to $0.01 m^2/s$ giving a very long decay time of 1.5 s.  
On the other hand, the diffusion coefficient of the ions induced by the collisions with neutral 
is about 
%$D_{\perp i} = \rho_{i}^2 / \tau_{i-n}
 $D_{\perp i} = 1 m^2 /s $. It is larger than $D_{\perp e}$ mainly 
due to the low magnetization of the ions. It gives a predicted decay time of 15 ms indeed very long
compared to the measured value.

In toroidal plasmas with sheared field lines, the trapping of particles inside banana orbits modifies
the diffusion coefficient if the trapping time is long compared to the collision time. 
This induces the so-called neo-classical effects. In our device,
it is easily shown that the ions are not efficiently trapped inside the magnetic mirrors due to the high collisionnality.
However, a large component of the distribution function of the electrons is actually trapped
during a sufficiently long time inside the banana orbits. In this situation, the neoclassical 
diffusion takes into account the poloidal Larmor radius and the local aspect ratio of the 
magnetic surface for the evaluation of the random-walk process of diffusion. This gives a neoclassical 
decay time about 10 times shorter than the decay time evaluated without trapping of the electrons
but still too long compared to the measured value.
In conclusion, the classical diffusion and the neoclassical diffusion for ions and electrons,
assuming ambipolar diffusion or not, are incompatible with the decay time
measured in the experiment.

Finally, the decay time has to be compared with the value predicted by the Bohm diffusion coefficient.
Assuming an electronic temperature of 4 eV, the Bohm diffusion coefficient is $D_{B}$ = 17 $m^2$/s
corresponding to a decay time of 1 ms calculated over a decay radius of 20 cm. This value is half the measured value.
If the decay time is calculated over a smaller radius, it is found in the range 0.1 to 0.5 ms.
It is important to note that an agreement with the measured value can be obtained if
the electron temperature is about 2 eV in the final stage of the decay. 
This point has to be investigated in next measurements.

Further investigations are needed to determine the mechanism of radial diffusion.
The temporal evolution of the electron temperature has to be measured during the plasma decay.
Moreover, different ways to estimate the radial transport have to be designed. 
In particular, the transport coefficient can be evaluated by following the time response to a density perturbation
in the central plasma, for instance a rapid change in the ionization source term.
 On the other hand, the very fast decay recorded just after switching-off the discharge 
is most probably determined by the very fast parallel collection of the energetic 
ionizing electrons. Finally, further investigations, especially
a precise mesurement of the temporal evolution 
of the electron energy distribution function would allow to check the diffusion mechanism. A plasma decay 
due mostly to volume recombination is highly improbable due to the low density of the plasma. The dependence 
of the decay time on collisionnality has to be investigated changing the gas pressure. 
To conclude, by comparing the final decay time of the plasma when the toroidal current is present or not, 
it is possible to assert that the confinement time is increased by more than a factor 10 when the poloidal 
magnetic field is established.

\section{CONCLUSION}
In summary, we have presented a new toroidal magnetized plasma laboratory device 
including a circular internal conductor establishing a stabilizing and confining poloidal magnetic field.
A high quality of the confinement is obtained and beyond a critical safety factor,
the plasma is found free of low-frequency instabilities. The radial transport is suspected 
to be in agreement with the Bohm diffusion, higher than the classical radial transport of electron and ions
even taking into account the partial trapping of the electrons inside internal magnetic mirrors.
This radial transport is considerably lower than the so-called anomalous radial transport 
observed in classical simple magnetized torii with circular field lines 
that are intrinsically turbulent laboratory  devices. 

The key parameter for the achievement of this stable toroidally confined plasma 
is the obtained low value of the radial electric field, but the
effective parameter is the profile of the safety factor. 
More precisely, the torsion of the magnetic field lines has to be
sufficient to produce an efficient mixing of the trajectories of the ionizing
and plasma electrons in order to get an equipotential volume of plasma.
\\

\section{ACKNOWLEDGMENTS}
The author is indebted to Prof. A. K. Sen and to G. Antar, I. Nanobashvili, E. Gravier, X. Caron, Y. Camenen for stimulating discussions. The author gratefully acknowledges L. Couedel for the careful correction of the manuscript. 
Expert technical assistance was provided by G. Vinconneau, A. Ajendouz and by the late Mr. K. Quotb.
This work has been partially supported by Centre National de la Recherche Scientifique (INSIS and Fédération de Recherche "Fusion par Confinement Magnétique"), and by European Union (European Regional Development Fund, ERDF).

\end{document}